# Connectivity in Social Networks

Sieteng Soh[1], Gongqi Lin[1], Subhash Kak[2]
[1]Curtin University, Perth, Australia
[2]Oklahoma State University, Stillwater, USA

**Abstract**
The value of a social network is generally determined by its size and the connectivity of its nodes. But since some of the nodes may be fake ones and others that are dormant, the question of validating the node counts by statistical tests becomes important. In this paper we propose the use of the Benford's distribution to check on the trustworthiness of the connectivity statistics. Our experiments using statistics of both symmetric and asymmetric networks show that when the accumulation processes are random, the convergence to Benford's law is significantly better, and therefore this fact can be used to distinguish between processes which are randomly generated and those with internal dependencies.

*Keywords:* Social networks, connectivity, value of a network, first digit phenomenon

## I. Introduction

The analysis of social networks from the perspective of their effectiveness as a medium of education, information diffusion, advertising and prediction [1],[2] is of great importance. In the naïve view, the value of a social network is correlated with its connectivity although what function of its size it should be has been much debated. Since in a communications network with $n$ nodes, each can make $(n-1)$ connections with other nodes, the total value of the network is proportional to $n(n-1)$, that is, roughly, $n^2$ (Metcalfe's Law). But a more careful analysis indicates that the potential value of a network of size $n$ grows in proportion to $n \log n$ [3] since in a social network one needs to take into account not only the members in it but also the local connectivity that is not uniform.

According to Zipf's Law, originally proposed to explain the linguistic phenomenon that the usage of words trails off in an harmonic fashion, the second element in the collection has about half the measure of the first one, the third one will be about one-third the measure of the first one, and so on. In general, in other words, the $k$ th-ranked item will measure about $1/k$ of the first one. It has been argued [3] that the over-valuation of communications companies based on the Metcalfe's Law was a contributing factor in the bursting of the Dot-Com bubble in 2000. Is it possible that the current valuation of social network companies may be likewise based on models that are not reliable?

In social networks the number of connections associated with each node is known, and the activity logs of each node is potentially available. But unlike a communication network defined in terms of each node that corresponds to physical hardware [4],[5], in a general social network, the problem is that many nodes may be fake for they are generated by software and there is no easy way to tell the fake ones apart from the real ones, and many nodes are inactive. Furthermore, analysis of their activity patterns is not easy given that the most popular social networks consist of hundreds of millions of nodes. The connectivity of each node cannot be considered a consequence of independent accumulation processes since there is also the social pressure on the participants to pad their number of friends.

Fortunately, it is possible for us to determine if the accumulation process is to be viewed as independent uniform variables. Given this we propose the use of Benford's Law in assessing how reliable is the connectivity data on social networks. According to Benford's Law the first digits of random accumulation processes are more common than larger ones [6]-[12]. The logic behind is that the accumulation process can stop at any number and since numbers with smaller first digits would have been traversed on the way to larger numbers, their probability is higher. This law is seen to describe the empirical data from a wide variety of sources such as electricity bills, population figures, and so on. The fact of the probability of occurrence of first digits is not uniform was first observed by Newcomb [6] and rediscovered by Benford over fifty years later [7]. Various other explanations such as scale invariance and counting explanations have also been advanced for this law.

The main contributions of this paper are twofold. First, it proposes the use of Benford's law to verify the validity of social network statistics for which the data has been used from well-regarded sources [14]-[16]. The use of Benford's Law has been proposed in forensics as in audit of numerical data [13]. To the best of our knowledge it has not been used in the study of social networks and we believe it can be put to good use there. Second, the paper presents experiments using the statistics of both asymmetric networks (for example, Twitter), and symmetric networks (for example, Facebook) to show the merit of using the law for such verification. The layout of the paper is as follows. Section II describes counting processes and Benford's law. Section III presents experimental methodology and our findings. Section IV concludes the paper.

## II. Counting Processes and Benford's Law

Counting processes may be assumed to have an underlying generator that produces uniform distribution of which arithmetical processes are a good example [17],[18]. A counting process may be considered to be uniformly distributed over the range $\{1, …, S\}$. For a very large number of such processes with random values of $S$, the set of numbers will satisfy Benford's Law and the leading digit $d$ ($d \in \{1, ..., 9\}$) occurs with probability

$$P(d) = log_{10}(1 + \frac{1}{d})$$

In other words, it is a characteristic of a mixture of uniform distributions [12]. Benford's Law may also be used to predict the distribution of first digits in other bases besides decimal. For any base $b \geq 2$, the general form is:

$$P(d) = log_b(1 + \frac{1}{d})$$

When $b = 2$, the probability of the first digit being 1 is trivially equal to 1. The analysis that yields the above law also can be extended to obtain the probability distribution for the second and subsequent digits. As is obvious, the distribution gets progressively less pronounced beyond the first digit. Table 1 gives the predicted frequencies for both the first and second digits.

Table 1. First and second digit frequencies

|           | 0     | 1     | 2     | 3     | 4     | 5     | 6     | 7     | 8     | 9     | Mean  |
|-----------|-------|-------|-------|-------|-------|-------|-------|-------|-------|-------|-------|
| 1st digit | -     | 0.301 | 0.176 | 0.125 | 0.097 | 0.079 | 0.067 | 0.058 | 0.051 | 0.046 | 3.441 |
| 2nd digit | 0.120 | 0.114 | 0.109 | 0.104 | 0.100 | 0.097 | 0.093 | 0.090 | 0.088 | 0.085 | 4.187 |

It would be correct to assume that if empirical data follows Benford's Law, it is generated by a mixture of independent uniform processes.

### III. Experiment Methodology

We evaluate our hypothesis using the Twitter and Facebook statistics that we obtained from Socialbakers [16] and the Tencent statistics from [2] that contain 6095 followers. The Socialbakers website on July 9, 2014 contained 486250 Twitter profiles each of which shows a pair (follower, following), with 'follower' value of 0 to 54353496, and the number of 'following' between 0 and 2428730. Our experiments consider both the 'follower' and 'following' statistics. For Facebook statistics, the website contained the Fans information of 3532700 pages, of which 3484364 pages had at least one fan.

**3.1. Twitter**

Let ER_S($N$) and ER_R($N$) respectively denote a sequence of $N$ Twitter profiles in sorted and random orders of total number of followers. Similarly, we use ING_S($N$) and ING_R($N$) to represent a sequence of $N$ Twitter profiles in sorted and random orders of total number of followings, respectively. To generate ER_S($N$) and ING_S($N$), we used the option provided in [16] to sort the profiles in order of followers and followings, respectively and selected the first $N$ profiles. Note that one can generate a sequence of random order of total number of followings, *i.e.*, ING_R($N$) from ER_S(486250); similar method can be used to obtain ER_R($N$ from ING_S(486250). However, to further refine the random order of selected $N$ profiles, we randomly selected ten sets of $N$ profiles for both 'following' and 'follower', and reported each result as the average of using the ten sets.

In the first experiment, we aim to see the effect of using sorted and random input on our hypothesis. Figure 1 shows the plot of the total number of profiles whose 'following' or 'follower' numbers start with digit 1, 2, …, 9 for ING_S(20000), ING_R(20000), ER_S(20000), and ER_R(20000); the figure also plot $P(d) = log_{10}(1 + \frac{1}{d})$, for $d = 1, 2, 3, …, 9$. Our results show that the plots of the four types of input are very close to that of $P(d)$, demonstrating their conformance to Benford's law. Further, both random input sets produce closer matching to the law as compared to their corresponding sorted input sets. Note that in our subsequence experiments we consider only random order of profiles.

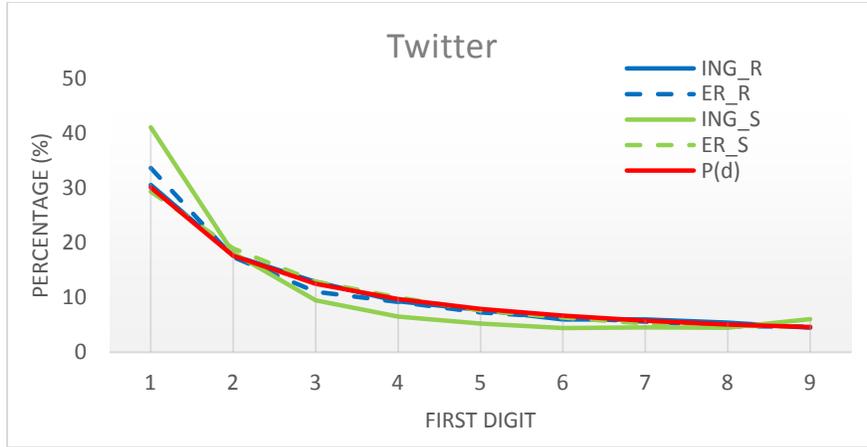

Figure 1. Sorted versus Random Profiles for 'Following' and 'Follower' (*N*=20000)

In the second experiment, we aim to see the effects of increasing total number of profiles on their conformance to the Benford's Law for both ING_R(*N*) and ER_R(*N*) with 5000, 10000, 15000, and 20000 profiles; each plot is the average over ten sets of data. As shown in Figure 2 (a), the average Twitter statistics for each of the four profile sets for 'following' closely matches the First Digit Law.

Further, we notice that there is an insignificant impact of the size of profile sets on the results. Figure 2(b) shows the results for Twitter followers, i.e., ER_R(*N*), for *N*=5000, 10000, 15000, and 20000. The figure shows that the statistics for followers are also very close to the First Digit Law, although they are not as close as their corresponding 'following' statistics, especially for digit '1'. Specifically, for this case, larger sized profile sets improve the results of conformance to the law.

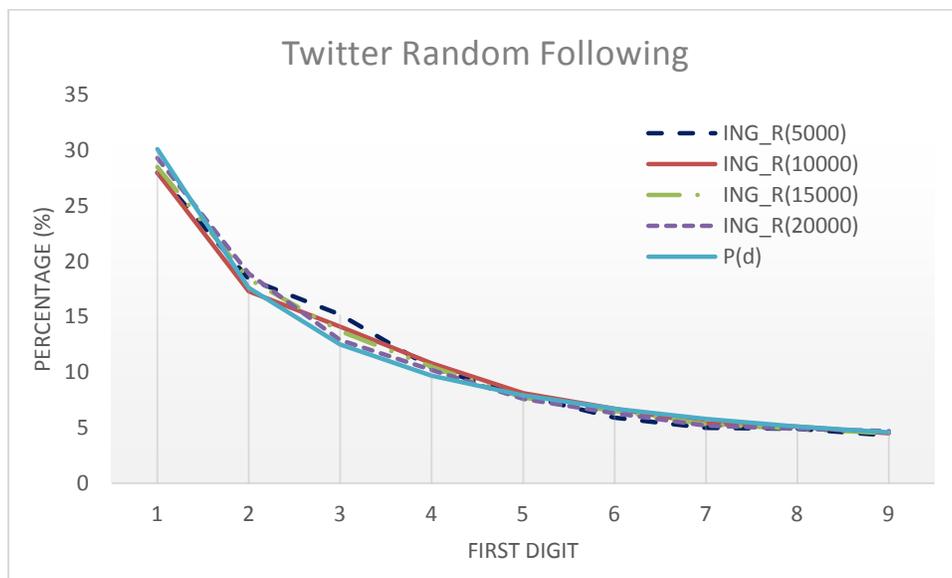

Figure 2 (a). Random 'Following'

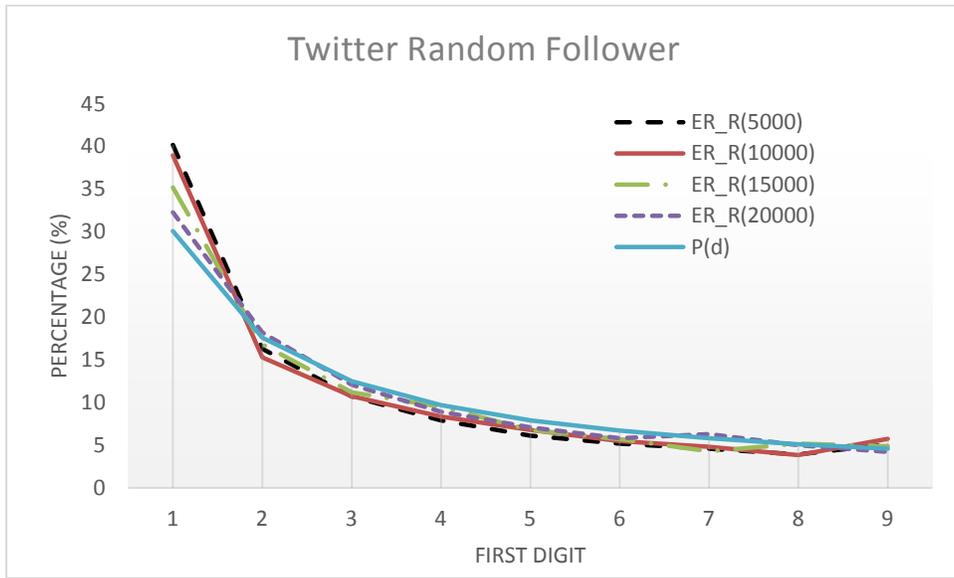

(b) Random 'Follower'

Figure 2. Results for Random 'Following' and 'Follower' (*N*=5000, 10000, 15000, 20000)

To further analyze the findings, we repeat the experiment for larger number of 'following' profiles, *i.e.*, *N*=30000, 40000, …, 100000, and *N*=486250. The results in Figure 3 further confirm the previous results.

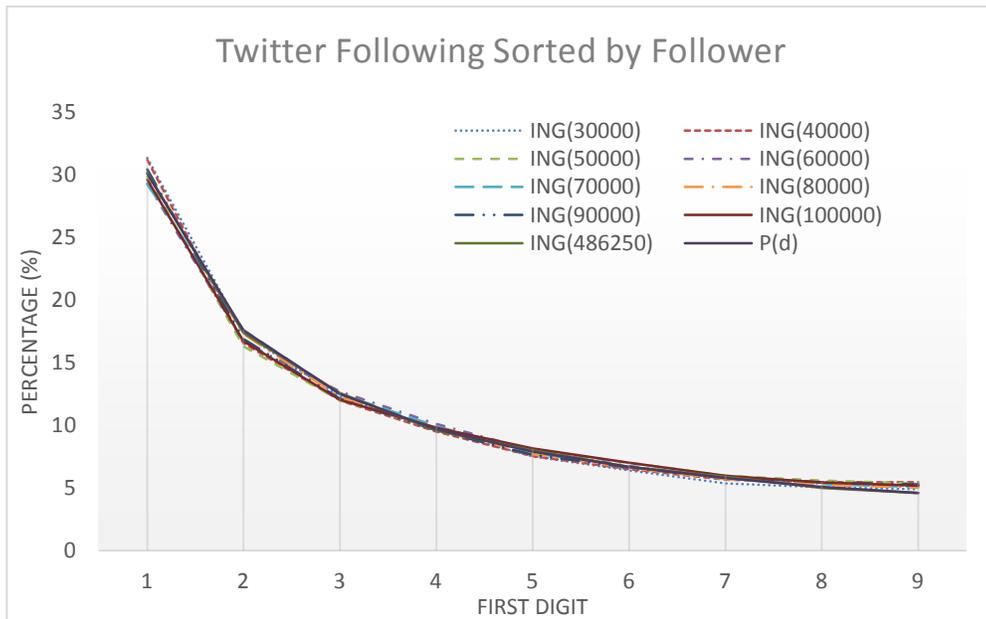

Figure 3. Twitter Random 'Following' (*N*=30000, 40000, …, 100000, and *N*=486250)

## 3.2. Results Using Other Social Media

In this section, we show our results on two other social media, *i.e.*, Tencent and Facebook. As shown in Figure 4, the statistics for Tencent followers (QQ) are very close to the First Digit Law. For Facebook statistics, we randomly selected *N*=20000 of the 3532700 pages and plotted their fans in Figure 5. The result in the figure shows random Facebook Fans also closely follow the Benford's law.

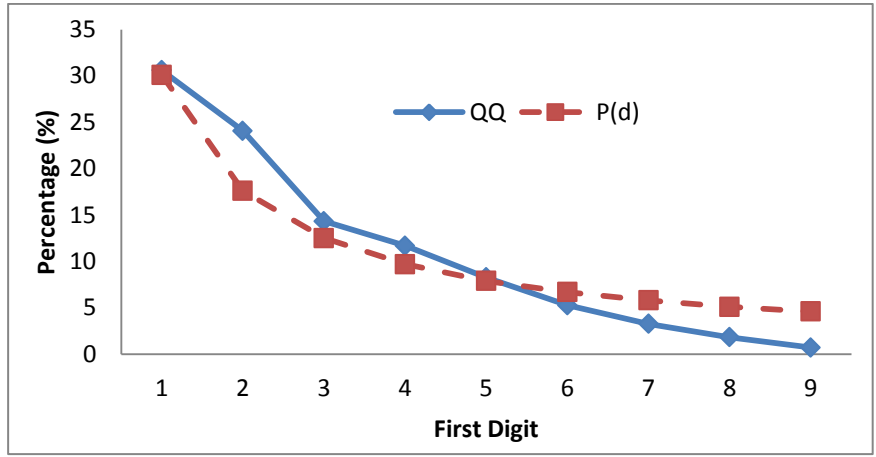

Figure 4. Tencent Random Follower (*N*=6095)

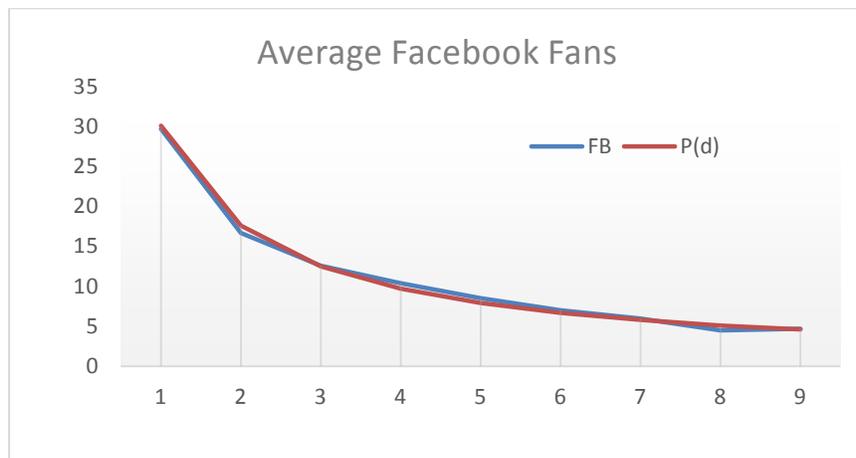

Figure 5. Facebook Random Fans (*N*=20000)

## IV. Conclusions

Social network connectivity [19]-[21] is changing society in unprecedented ways and it has created new challenges related to integrity of data as well as that of security [22]-[25]. For the latter also, the use of Benford Law related techniques of analysis can assist in the detection of intrusion and other attacks, and this will be discussed elsewhere.

This paper has shown the application of the Benford's Law to examine the connectivity data in social networks. Our results using random statistics of Twitter, Facebook, and Tencent show their conformance to the law. Non-conformance to the First Digit Law may be due to several reasons. In symmetric networks, *e.g.*, Facebook, the individual members may try to pad their list of friends for purposes of bragging. For this case, we believe that the connectivity statistics in asymmetric networks like Twitter are more reflective of independent variable hypothesis than the connectivity statistics in symmetric networks. The departure from Benford's Law may also be due to fake accounts. According to one source, about 25% of Facebook accounts are fake [14]. Likewise, the Twitter accounts of celebrities have many fake follower accounts that have been created to make the celebrities more popular than they actually are [15]. If forensics based on Benford's Law make it possible to determine how consistent connectivity data on social networks are, that would be of significant value.